# Mysterious frequency combs in erbium-doped fiber lasers at low temperatures

**Alexander Bekker, Gefen Levite, and Baruch Fischer**

The Andrew & Erna Viterbi Faculty of Electrical and Computer Engineering, Technion, Haifa 32000, Israel

E-mail: fischer@ee.technion.ac.il

**We report on the observation of puzzling nonuniform, but ordered, frequency combs in erbium-doped fiber lasers at low temperatures, between 3 *K* and 14 *K*. At ~14 *K* the combs disappear. The combs contain ~240 lines and have nonuniform frequency spacings that starts from ~25 GHz at ~1540.5 *nm* and decreases, almost monotonically, to ~100 MHz within (1.8-2) *nm*. We discuss possibilities that they result from enlarged regular mode-comb spacings, from prime numbers-based solitary waves free of four-wave mixing or result from gain gratings. However, we think that the combs originate from multi-lasing lines allowed in inhomogeneous gain broadening, dominant at low temperatures, together with a small power dependent homogeneous broadening part, responsible for the spacing between the lines.**

Laser frequency combs have become an important base for many purposes, such as spectroscopy and metrology[1-6]. They result from longitudinal cavity modes that have an equal frequency spacing. The combs are based on longitudinal modes of a laser cavity and usually have a close to equal spacing between the modes. In our case the mode spacing is $\approx 5\,MHz$, corresponding to a 40 *m* ring fiber cavity length. Very rarely a mode-comb was also produced with a nonuniform spacing[7]. In this work we observed mysterious nonuniform frequency combs of lines in a pumped erbium-doped fiber (EDF) laser at low temperatures, between 3 *K* and 14 *K*. They contained ~240 lines that started from a wavelength 1540.7 *nm* with a spacing of ~25 GHz (0.2 *nm*) that almost monotonically decreased to ~100 MHz (~0.0008 *nm*) at ~1542.5 *nm*. (There were denser lines, but they were week and hard to be measured.) The above values correspond to the first (top) comb example in Fig. 1a, while the others have slightly different values. The combs were seen within a small wavelength interval of (1.8-2) *nm* and they disappeared above ~14 *K*. The comb evolution is shown in Fig.1b. We excluded a few explanations, including one that consider their spacing as an enhancement of the laser mode spacing by 3 orders of magnitude, or originating from gain gratings[8]. We also thought about solitary or fixed-point lasing waves that follow the prime numbers sequence (suggested by Eyal Buks) that fundamentally eliminates four-wave mixing (4WM). It provided a good fit to the experimental decreasing spacing, but has other problems, as we explain below. We think however that most likely the comb originated from many lasing lines allowed due to inhomogeneous gain broadening[9-15] that is dominant at low *T*, together with a small homogeneous broadening part that was responsible for the spacing between the lines. The decreasing power of those lines with the homogeneous power dependent width were responsible for the decreasing spacing. The combs disappearance at about 14*K* resulted from a change from an inhomogeneous broadening regime to a homogeneous broadening dominance that mostly causes a single wavelength lasing. See the combs *T* dependence in Video 1 in the Supplementary part.

     In the experiment we used a closed cycle cryostat to cool down an EDF of length 20 *m*. The overall fiber ring cavity length was 40 *m*. The EDF was a single mode fiber with a core diameter of $5\,\mu m$. A laser diode with a wavelength of $980\,nm$ was used to pump the EDF. The other side of the cooled EDF was connected to an optical spectrum analyzer (OSA) by a 10% output coupler. We placed in the ring cavity two 50 *dB* isolators that made the light propagation one-directional.

Near a wavelength of $\lambda \approx 1540.5\ nm$ and below ~14 K, the measured optical spectrum revealed an ordered nonuniform comb exhibited by narrow peaks at a sequence of wavelengths denoted by $\lambda_n$, where $n = 1, 2, 3…$, starting from the left side with the lowest comb wavelength, as shown in Fig. 1a. We emphasize that the combs spectra, the values $\lambda_n$ (including the starting wavelength $\lambda_1$), the spacings $\Delta\lambda_n = \lambda_{n+1} - \lambda_n$ and the powers $P_n$ slightly varied from experiment to experiment and when the operating parameters, especially the pumping level, changed. Besides the spectra in Figs 1 and 2 we measured many other spectra at various temperatures and pumping levels. The OSA measures less than 10% of the light in the fiber. The pumping was raised from zero by increasing the current to the pump diode-laser. The comb appeared as the threshold current ~26.7 mA was passed, as shown in Fig. 1b and in video 2 at in the Supplementary Information part. We started with a temperature of 3 K measured at the copper reel that surrounded the fiber and kept that along this experiment. Nevertheless, the temperature inside the fiber presumably could raise due to the pumping. A close view of the EDF comb and the gain at low T is given in Figs. 2a and 2b. Fig. 2c shows a broader spectrum from low T until a room temperature, taken by a low resolution OSA. The general tendency is of having the gain region at lower wavelengths as the temperature was lowered.

There can be a few explanations to the combs. We start with the one that we prefer and then give the other possibilities and discuss why we reject them. We think that the combs were composed of multi-line lasing allowed in inhomogeneous broadening gain media, while the homogeneous part was responsible for the spacing. We see from the measurements in Figs. 1a and 2, confirmed below by Figs. 3c and 3d based on many data points, that

$$\begin{aligned} P_n / \Delta\lambda_n &\approx\ \text{Constant} \approx\ 0.5\ mW/nm \\ \Delta\lambda_n &\approx 2 P_n \quad (\Delta\lambda_n\ \text{in}\ nm,\ \text{and}\ P\ \text{in}\ mW) \end{aligned} \quad (1)$$

where $P_n$ is the *n-th* line power in *mW*, and $\Delta\lambda_n = \lambda_{n+1} - \lambda_n$ is the *n-th* wavelength spacing in *nm*. As said above, the spacing is associated with the homogeneous broadening. A minor issue can be on which spacing to take for $P_n$, the next one, as we did, the former one, or an average of both. We also note that we don't know exactly the power in the fiber core. The constant in Eq. 1 was roughly estimated by taking for the first line a power inside the fiber of $P_1 \approx -10 dBm = 0.1\ mW$ (an output coupler leads only 10% of the light power in the fiber to the OSA), and for the first spacing $(\lambda_2 - \lambda_1) \approx 0.2\,nm$.

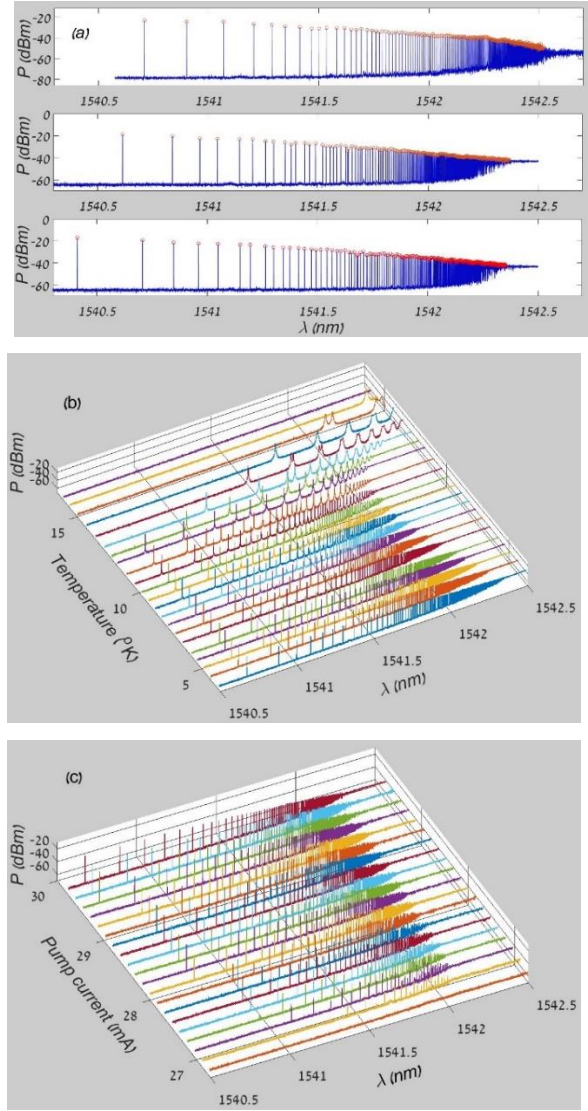

**Fig. 1. Frequency combs**: (a) Typical spectra at various temperatures between 3 *K* and 14 *K* and various pumping. The top comb starts at a wavelength $\lambda \approx 1540.7\,nm$ with a spacing of $\approx 0.2\,nm$ and ends at $\approx 1542.52\,nm$ where a transition occurs to a regular cavity mode spectrum. Only 10% of the light power in the fiber was directed to the OSA by an output coupler. (b) Comb as a function of temperature with a pumping current of 35 mA : See also video 1 in the Supplementary Information part. (c) Comb lines evolution with pumping at 3.7 *K*. The pumping was done by a current injection to the pump laser-diode in steps of 0.16 *mA*. The threshold was at ~26.7 *mA*. See also video 2 in the Supplementary Information part.

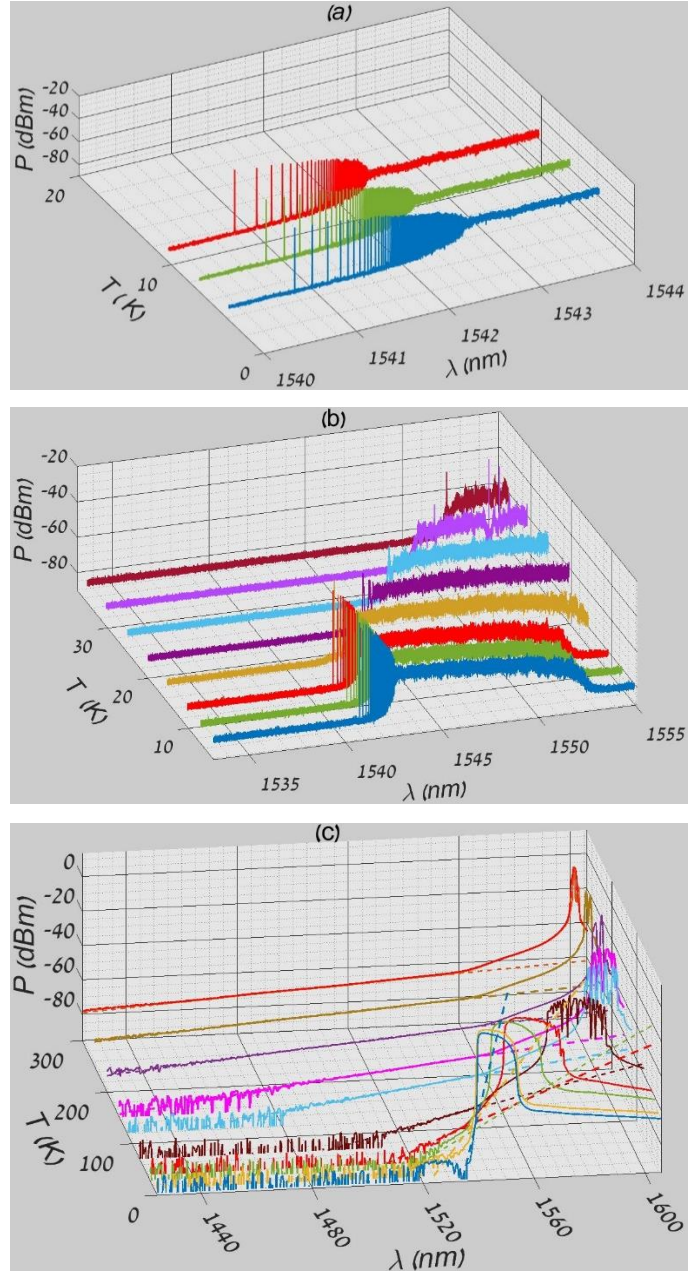

**Fig. 2. Typical spectra of the combs and the gain at various temperatures and ranges:** (a) Focus on the combs region showing high resolution spectra for three temperatures between 3 $K$ and 14 $K$. The lower one starts at about 1540.8 $nm$ and ends at $\lambda \approx 1542.8\,nm$ (an interval of $\approx 2\,nm$). (b) A broader view of the combs and the gain spectra for various $T$. They show three transitions: Two relate to the combs sharp start and end. The third one is of the whole gain region that ends at $\approx 1553\,nm$. (c) An overall view of the spectra for a large temperature range, from low $T$ to a room temperature (taken with a low resolution OSA). The dashed straight lines have slopes that correspond to and follow Bose-Einstein distribution for each $T$.

The experimental measurements shown in Fig. 3 and especially $\Delta\lambda$ vs. $P$ in a linear scale in Fig. 3c and in a log-log scale in Fig. 3d show the right slope that fits the relation $\Delta\lambda_n \approx 2P_n$ nm ($P$ is in *mW*). The log scale that spreads the data over a few orders of magnitude helps to see the low values. As noted above, these values correspond to the first (top) comb example in Fig. 1, while the others have slightly different values, but Eq. 1 approximately holds to them as well. Then, here and in other experiments like in hole burning, we conclude that for homogeneous broadening $\Delta\lambda \sim 2P$ nm ($P$ in *mW*) at least for the low $T$ and $P$ regimes. It means that the spacing or the hole (in hole burning) is proportional to the line power and here the line power per wavelength (or frequency) is almost constant. When the power decreased the spacing decreased accordingly. The spacing and so the constant changed as the overall pumping and the power changed due to the small homogeneous broadening that still exists at the low $T$ regime. Therefore, we can explain the decreasing spacing by the decreasing $P_n$. A more basic formula for Eq. 1 would replace the power $P$ by the power density $I \approx P/(0.8\pi r_f^2) = (P/15.68)$ $mW/(\mu m)^2$ in the fiber (where we took $r_f = (5/2)\mu m$ for the fiber-core radius and a filling factor of 0.8). Then $\Delta\lambda \approx 31I$ $nm$, where $I$ is given in $mW/(\mu m)^2$.

The nonuniform combs line powers $P_n$ depend on the gain and loss curve. After pumping beyond the threshold most lines that have the same power per wavelength-spacing (Eq. 1) start to appear. (See Fig. 1b and video 2 in the Supplementary Information part done at *T*=3.7 *K*.

. The spacing followed the intensity and became small as well. The final spectra for high pumping rates in Fig. 1a with the decrease of the power with *n* followed the slightly decreasing gain-loss pattern. It doesn't correspond to the line appearance order that was not at the first and highest power line, but more to the low power dense lines (see Fig. 1b, video 2 in the Supplementary part and a note above), but most of the lines appeared quite quickly after passing the threshold. At higher pumping level, the power $P_n$ almost monotonically decreases, and consequently the line spacing, that results from the homogeneous broadening part, decreases as well.

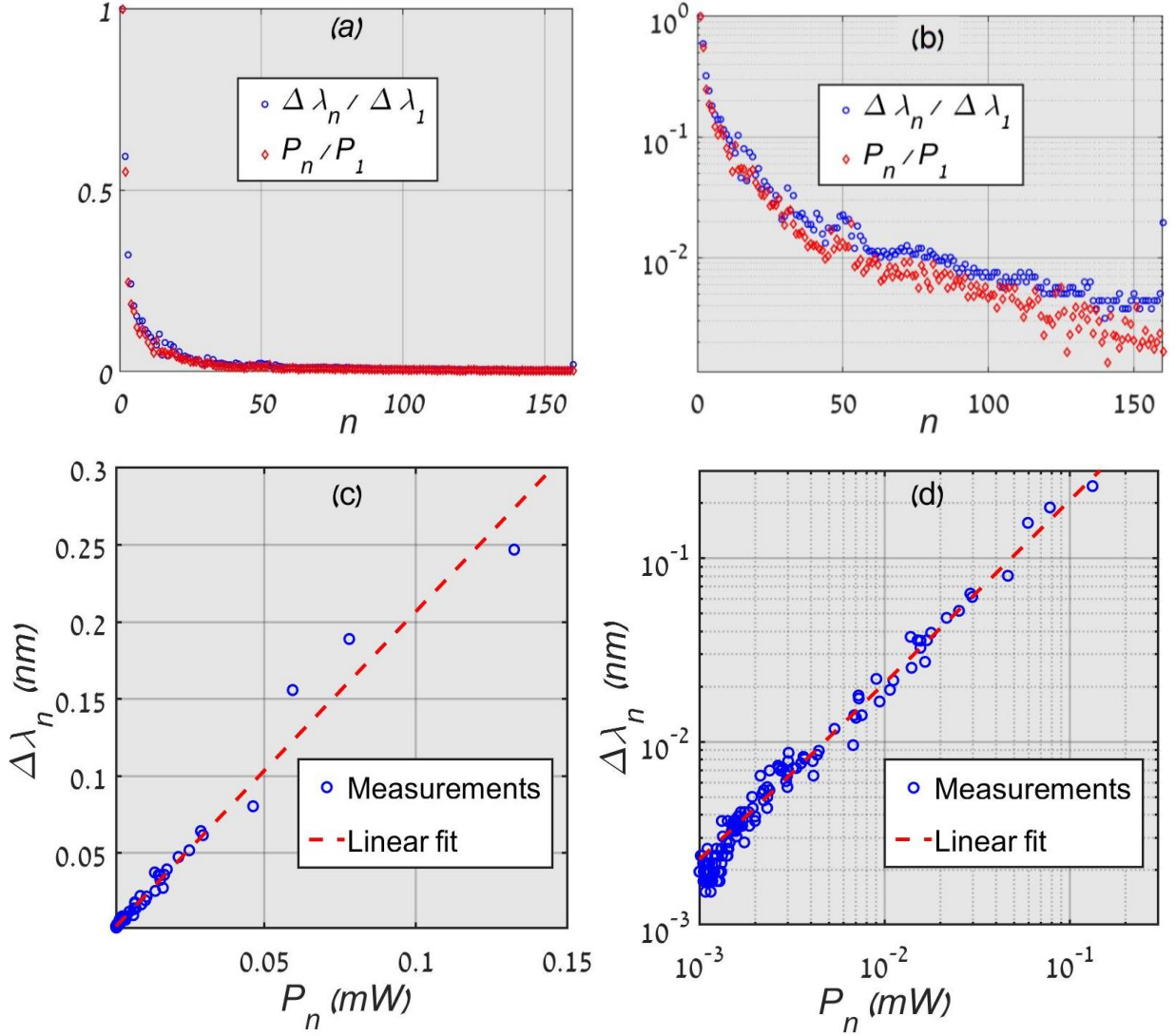

**Fig. 3. Power and spacing:** (a) and (b) Normalized power and spacing vs. $n$ (150 out of ~240 lines) in linear and semi-log scales. (c) and (d) $\lambda_n$ vs. $P_n$ of a specific spectrum in a linear and log-log scales. The log scale spreads the data over a few orders of magnitude and helps to see the low values. The linear fits in (c) and (d) (dashed red lines) follow Eq. 1 giving in the linear scale in (c) a slope of slightly larger than 2, and in the log-log scale in (d) a slope of 1 and a shift in the y-axis direction of slightly larger than $\log 2 \approx 0.3$.

The decrease of the lasing power lines $P_n$ with $n$ can be explained by the decreasing gain-loss profile and the spacing $\Delta\lambda_n$ decreases with $n$ accordingly, as explained in and after Eq. 1. The spacing results from the range of the small homogeneity that exists at low $T$. While the overall pattern is of decreasing spacing values with *n*, there are a few exceptions of certain lines with increased spacing compared to their former lines, as seen in Figs. 1 and 2. They usually correspond to a larger power compared to the former (or/and the following) line.

The first spacing in the top comb of Fig. 1, $(\lambda_2 - \lambda_1) \approx 0.2 \ nm$, corresponds to $(\nu_2 - \nu_1) \approx 3.1 \times 10^{10} \ Hz$. It is close to what was found in several former papers obtained from hole burning experiments[10-13] and resonant fluorescence line narrowing measurements[14,15] giving for the homogeneous broadening at low *T* an order of $(10^{10} - 10^{11}) \ Hz$. For example, a hole burning experiment at 4 *K* yielded[10] a homogeneous width of 0.21 *nm*. Here we add to it, what we already noted, that the homogeneous width depends on power according to $\Delta\lambda \approx 2P \ nm$ (*P* in *mW*) at the low *T* and *P* regimes. Therefore, the spacing and the homogeneous broadening width strongly depend on the power *P* and on *T*. As mentioned above, this result can explain the decreasing spacing of the lines as *n* increased. We also note that there was a gradual transition at about 14 *K* to a homogeneous broadening dominance over inhomogeneity that mostly causes a single line lasing.

We can summarize the explanation that the lasing combs at low *T* below 14 *K* was possible due to the dominance of inhomogeneous broadening that when the pumping raised allows multi-lasing lines and the spacing resulted from the relatively small homogeneous broadening part. While in the high *T* case there is mostly one lasing line that depresses the surrounding gain, the low *T* case allows many lasing lines, and the lasing combs (compared to the regular frequency combs composed of longitudinal modes and can be seen in a single lasing line that includes many longitudinal modes.)

We also checked and excluded a few other explanations. The first one was that the lines were gain affected enlarged longitudinal modes of the cavity. We note that the regular cavity mode frequency spacing of the comb was ~5 MHz. It would be therefore an enhancement of more than three orders of magnitude to ~25 GHz. We changed the cavity length, but the comb didn't change accordingly. We also thought that the sharp gain growth close to this region may be responsible for the enhanced spacing due to the Kramers-Kronig relation that may give the needed refractive-index change. It would require a reduction of the index to be close to zero. However, the analysis showed that it is unrealistic. Gain gratings[8] that change the index and a small Fabry-Perot that can form large spacing combs were raised as well and rejected. We also thought about an hypothesis with a new kind of solitary waves that follow the prime numbers series, explained below. It determines the lines and produces a very good fit to the experimental accumulated spacings $\lambda_n - \lambda_1$. Then, the power $P_n$ can be derived from the spacing $\Delta\lambda_n$ according to Eq. 1. It is in the same way, but in reverse to our preferred explanation that the combs result from multi-lasing lines due to inhomogeneous and homogeneous broadening. There $\Delta\lambda_n$ was derived from $P_n$, and here $P_n$ from $\Delta\lambda_n$. The primes hypothesis, however, failed to explain the exact spacing and the line locations, and it needs a justification to the log relation. This hypothesis is explained below.

We finally note about Fig. 2c that a significant part of the measurable spectra (a large portion was below the detectable power, especially for low *T*) of wavelengths below the lasing region are straight-lines with slopes, shown by the dashed lines in the Fig. 2c where the power is in *dBm*, that fit Bose-Einstein distribution with the corresponding temperature. It alludes to the possibility that the photon system was there in thermal equilibrium that can lead to Bose-Einstein condensation[16-18].

We elaborate on the prime numbers hypothesis that was raised by Eyal Buks. In this proposition the frequency $\nu_n$ is expressed as,

$$(\nu_n - \nu_1) = \Delta\nu_0 \ln x_n, \qquad (2)$$

where $\Delta\nu_0$ is a constant. The frequency $\nu_n$ is associated with the wavelength $\nu_n$ by $\nu_n \lambda_n = c/n_r$, where *c* is the speed of light in vacuum and $n_r$ is the refractive index of the silica fiber core. Consider the case where $x_n = p_n$ are prime numbers ($p_n = 1, 2, 3, 5, 7, 11, 13...$). Then a comparison between the experimentally measured and calculated sequences of $\nu_n$ shows a good agreement. Why is it expressed with a log? The main answer is that it fits the experiment but there are statistical reasons to justify that. If so, we can explain this proposition as we argue below.

If the log relation is accepted, we suggest the following explanation: Consider a 4WM process in which two photons with frequencies $\nu_i$ and $\nu_j$ are annihilated and forms two photons at frequencies $\nu_k$ and $\nu_l$ where $\nu_i + \nu_j = \nu_k + \nu_l$ (energy conservation). The phase matching condition (regarding $\vec{k}$ or momentum conservation) in 4WM is not a problem here because of the relatively small frequency differences and the short length (40 *m*) of copropagating waves. Due to the log properties, $\nu_i + \nu_j = \nu_k + \nu_l$ is translated to $x_i x_j = x_k x_l$. When the sequence $\{x_n\}$ is chosen according to the above equation (primes), there isn't any possibility for 4WM of a frequency pair to produce other two frequencies. This means that the prime series generates lines that fundamentally cannot undergo a 4WM process. They are free of 4WM and therefore cannot lose energy to other frequencies by 4WM. This makes them different from other frequencies that can lose power to any pair and form broad and shallow spectra. They are a kind of unperturbed solitary waves (fixed-point waves). The 4WM itself can be weak but it can be enhanced and amplified by the gain. In this hypothesis the power $P_n$ can be derived from the spacing according to Eq. 1, reversely from our preferred explanation where the spacing was derived from $P_n$. However, we rejected this explanation despite the plausible agreement with the experiment, especially of the accumulated spacing $\lambda_n - \lambda_1$, since we found that the spacing of a few first and other comb lines and the ratios of spacings (that eliminate the constant $\Delta v_0$) didn't follow exactly this theory. It is also difficult to find explanation to the log relation, to the starting line frequency $v_1$ and to $\Delta v_0$ (maybe it has to do with the inhomogeneous broadening width). In addition, the strength of the 4WM process is weak in the relatively short fiber despite the active gain environment. Also, the edf gain can contribute to wave amplification but not to 4WM nonlinearity with frequency differences larger than ~1 kHz. We nevertheless brought this theory because of its very good fit to the experimental spacing accumulation plot and its elegance that might be meaningful in other cases, such as in supercontinuum generation[19] or in many-body statistics systems at a finite temperature or noise that have a log relation like in the free energy[20-22].

**Acknowledgments:** This research was supported by the Israel Science Foundation.

**Author contributions:** All authors took part of the work. The initiation, ideas and the theory were provided by B. F and A. B. The experiment was done in the low temperature cryogenic system of Dr. Eyal Buks who participated in the experimental work. The fiber and optics parts and sides were prepared and conducted by A. B. and G. L. Eyal Buks didn't want to be an author of this paper, despite our willing to share with him the outcome and the paper, since he believed in another theory based on four-wave mixing and prime numbers, that we thought is incorrect in our experiment.

**Additional information:** Correspondence and requests for materials should be addressed to B.F. (fischer@ee.technion.ac.il )

## Supplementary Material:

**Video 1:** The comb *T* dependence from 3.6275 *K* to 16.4629 *K*. The comb disappears above ~14-15 *K*. The current of the pumping laser-diode was 35 *mA*.

spestra_vs_T_5fps.mp4

**Video 2:** The modes appearance as the current of the pumping laser-diode was increased from below to above the edfl threshold in steps of 0.016 *mA*, at *T*=3.7 *K*.

peaks5fps_PC2205_DE4.mp4